%% file: MAIN.tex
\documentclass{IEEEtaes}

\usepackage{color,array,amsthm}
\usepackage{graphicx}
\usepackage[caption=false, font=footnotesize]{subfig}
\usepackage{rotating}
\usepackage{pdflscape}
\usepackage{booktabs}
\usepackage{siunitx}
\usepackage{mathtools}
\usepackage{cuted}
\usepackage{multirow}
\usepackage{eqnarray}
\usepackage{amsmath}
\usepackage{placeins}
\usepackage{float}
\usepackage{lipsum}
\usepackage{flushend}
\usepackage{balance}
\usepackage{scalerel}
\usepackage{stfloats}
\usepackage{bigints}
\usepackage{textcomp}
\usepackage{tabularx}
\usepackage{cleveref}
\usepackage{booktabs}

\usepackage{makecell}
\usepackage{siunitx}
\usepackage{tcolorbox}

\usepackage{stackengine}

\usepackage[export]{adjustbox}

\usepackage{amsmath,amssymb,amsfonts,cite,graphicx,xcolor,color,multirow,array,eqnarray}
\usepackage{kantlipsum,lipsum,mathtools,tabularx,threeparttable,adjustbox,pdflscape,longtable,bigints}
\usepackage{flushend,booktabs,siunitx,textcomp,placeins,scalerel,cuted,breqn,csquotes,mwe,rotating,url}
\usepackage{algpseudocode}
\usepackage{algorithm}
\usepackage{setspace} 
\usepackage{nomencl}
\usepackage{etoolbox}
\makenomenclature

\usepackage{acronym}
\usepackage{array}
\begin{document}

\title{Modeling Lunar Power Harvesting and Battery Recharging In The Presence of Random Pointing Error} 

\author{BARIŞ DÖNMEZ}
\member{Graduate member, IEEE}
\affil{Polytechnique Montréal, Québec, CA} 

\author{YANNI JIWAN-MERCIER}
\affil{Polytechnique Montréal, Québec, CA}

\author{JULIEN MAGNÉ}
\affil{Volta Space Technologies, Québec, CA}

\author{GUILLAUME BLANCHETTE}
\affil{Volta Space Technologies, Québec, CA}

\author{PAOLO PINO}
\affil{Volta Space Technologies, Québec, CA}

\author{SÉBASTIEN LORANGER}
\member{Member, IEEE}
\affil{Polytechnique Montréal, Québec, CA}

\author{GÜNEŞ KARABULUT KURT}
\member{Senior member, IEEE}
\affil{Polytechnique Montréal, Québec, CA}



\authoraddress{Barış Dönmez, Yanni Jiwan-Mercier, Sébastien Loranger, and Güneş Karabulut Kurt are with Poly-Grames Research Centre, Department of Electrical Engineering, Polytechnique Montréal, and
Space Resources and Infrastructure Engineering Research Unit (ASTROLITH), Polytechnique Montréal, Québec, H3T 0A3, Canada. Julien Magné, Guillaume Blanchette, and Paolo Pino are with Volta Space Technologies, Québec, CA.
(e-mail: \href{mailto:baris.donmez@polymtl.ca}{baris.donmez@polymtl.ca}). 
}


\maketitle

\begin{abstract}
In our work, we establish a lunar power beaming environment between a low-lunar orbiter and a circular solar panel mounted on top of a lunar vehicle located at the Lunar South Pole. Due to the random nature of the pointing errors, various upper bounds reflecting the severity of these errors are considered. Although harvested power is modelled stochastically in the literature, we extend this randomness to battery-storage analysis by deriving PDFs of state of charge, accumulated battery capacity and energy, and end-to-end efficiency for both battery charging modes. Then, our derivations are verified with extensive Monte Carlo simulations. Hence, our derivations provide a mathematical model that enables prediction and helps us understand the randomness of the corresponding variables. For the given pointing error levels, we quantify the differences between charging modes in these battery metrics. Our work explains the stochastic nature of these battery metrics in a realistic lunar environment. Regarding the end-to-end efficiency metric, which is the ratio of the stored electrical energy to the transmitted electrical energy, we observed that slow-charging efficiency is immune to pointing errors, and this mode can utilize only one-sixth of what the link can deliver under a mild pointing error.
\end{abstract}


\nomenclature{$I(r,z)$}{Gaussian beam irradiance}
\nomenclature{$I_{0}$}{Maximum irradiance at the Gaussian beam center}
\nomenclature{$r$}{Radial distance from the centre of the Gaussian beam footprint}
\nomenclature{$z$}{Gaussian beam path length between the satellite and the solar panel}
\nomenclature{$w(z)$}{Gaussian beam radius at distance $z$}
\nomenclature{$P_{t,o}$}{Optical transmitted power of a Gaussian beam from the satellite}
\nomenclature{$w_{0}$}{Gaussian beam waist}
\nomenclature{$\theta$}{Gaussian beam divergence angle}
\nomenclature{$\lambda$}{Optical wavelength}
\nomenclature{$D_{t}$}{Satellite aperture diameter}
\nomenclature{$P_{c}$}{Collected power of the Gaussian beam}
\nomenclature{$A'$}{Receiver aperture area}
\nomenclature{$b$}{Receiver aperture radius}
\nomenclature{$v$}{Beam footprint spatial displacement error}
\nomenclature{$P_{a}$}{Analytical approximation of $P_c$}
%
\nomenclature{$\sigma_{\varphi_c}$}{Standard deviation of Gaussian pointing error}
\nomenclature{$\varphi_{c}$}{Component of the radial error angle RV}
\nomenclature{$\varphi$}{Radial error angle RV}
\nomenclature{$\sigma_{v}$}{Standard deviation of spatial displacement error}
\nomenclature{$P_{t_e}$}{Electrical input transmit power}
\nomenclature{$\eta_h$}{Optical-electrical PCE}
%
\nomenclature{$SoC$}{State-of-charge}
\nomenclature{$SoC_{0}$}{Initial SoC}
\nomenclature{$I_b$}{Battery recharging current}
\nomenclature{$C_b$}{Battery capacity}
\nomenclature{$T_{fc}$}{Fast charging total duration}
\nomenclature{$SoC_{FS}$}{Fast-to-Slow charging mode transition SoC}
\nomenclature{$\tau$}{RC time constant}
\nomenclature{$V_b$}{Nominal battery voltage}
\nomenclature{$T_b$}{Beaming time interval}
\nomenclature{$f_{X}(x)$}{Probability density function}
\vspace{1 cm}
\printnomenclature[1.5 cm] 

\section{Introduction}
\input{Sections/Introduction}

\label{Sec:intro}
\section{Lunar Power Beaming and Storage Model}
\input{Sections/SystemMod}

\label{Sec:sysmod}
\vspace{-0.4 cm}
\section{Results and Discussion}
\input{Sections/ResultsDiscussion}

\label{Sec:results}
\vspace{-0.5 cm}
\section{Conclusion}
\input{Sections/Conclusion}
\label{Sec:conc}
\section*{Acknowledgment}
This work is supported by Volta Space Technologies, the Consortium for Research and Innovation in Aerospace in Québec (CRIAQ), and the Natural Sciences and Engineering Research Council of Canada (NSERC).

\bibliographystyle{IEEEtran}
\bibliography{References}
\vspace{-0.25 cm}

\input{Biography}
\label{Sec:bio}

\end{document}

%% file: Sections/Introduction.tex
T{\scshape he} lunar-based studies have been gaining traction as the Moon offers a plethora of opportunities. Since the Moon is Earth's closest satellite, the Moon Village Association (MVA), comprising hundreds of attendees from space agencies, academia, and related fields, has been working on a permanent lunar settlement\cite{moonvillage, general1}. Furthermore, space vehicles can stop on the Moon for refuelling and maintenance before departing for interplanetary missions \cite{coggins2024nasa, orbittypes}. In addition, many entrepreneurs may be enticed by the prospect of a myriad of valuable materials beneath the lunar crusts, which can be excavated by lunar vehicles operating without interruption, owing to power solutions \cite{lunarmining, donmez2026multi}.


Power transmission in the space environment is an arduous task since the distances are in hundreds of kilometres; this is also valid for the Moon, as there is no existing power infrastructure \cite{donmezatp, donmez2025hybridfsorflunar}. On the other hand, ubiquitous radiofrequency (RF) and free-space optics (FSO) technologies can be used to establish wireless power transmission (WPT) in space \cite{lasertypes}. Each approach has advantages and disadvantages depending on the desired scenarios \cite{EHwindow}. Primarily, an optical laser transmitter sends a collimated beam, which significantly reduces path loss, as does the beam spot size, whereas a microwave signal spreads widely and hence can reach multiple locations at the expense of a high path loss \cite{lasertypes, ccetin2025advancing}. In RF transmission systems, the equipment is larger and heavier but offers higher power conversion efficiencies (PCEs) than FSO transmission systems \cite{marcinkowski2023lunar,donmez2025hybridfsorflunar}. A critical drawback of laser transmission is the misalignment errors \cite{donmezatp,donmez2024continuous}, particularly for a spacecraft orbiting around the Moon at a speed of kilometres per second \cite{donmez2025modeling}. Aside from the intrinsic bottlenecks, lunar dust can deteriorate the power reception and hence the communication links \cite{naqbi2024opticalpowerbeaminglunar, jiwan2025diffraction}. 

There are numerous lunar orbit types with varying critical parameters, such as altitude, period, inclination, and stability \cite{orbittypes}. Distant retrograde orbit (DRO), Earth-Moon Lagrange point (EMLP)-2 halo orbit, near rectilinear orbit (NRO), elliptical lunar orbit (ELO), frozen lunar orbit (FLO), prograde circular orbit (PCO), and low lunar orbit (LLO) have altitudes reaching around 70,000, 60,000, 75,000, 10,000, 8,800, 5,000, and 100 km, respectively. The LLO period is approximately 2 hours, whereas orbiting around DRO and EMLP-2 can take up to 14 days. The inclination of the DRO and ELO is suitable for equatorial coverage of the Moon; however, NRO provides coverage for the poles.

\begin{figure}[t!]
\centering
\includegraphics[width=1\columnwidth]{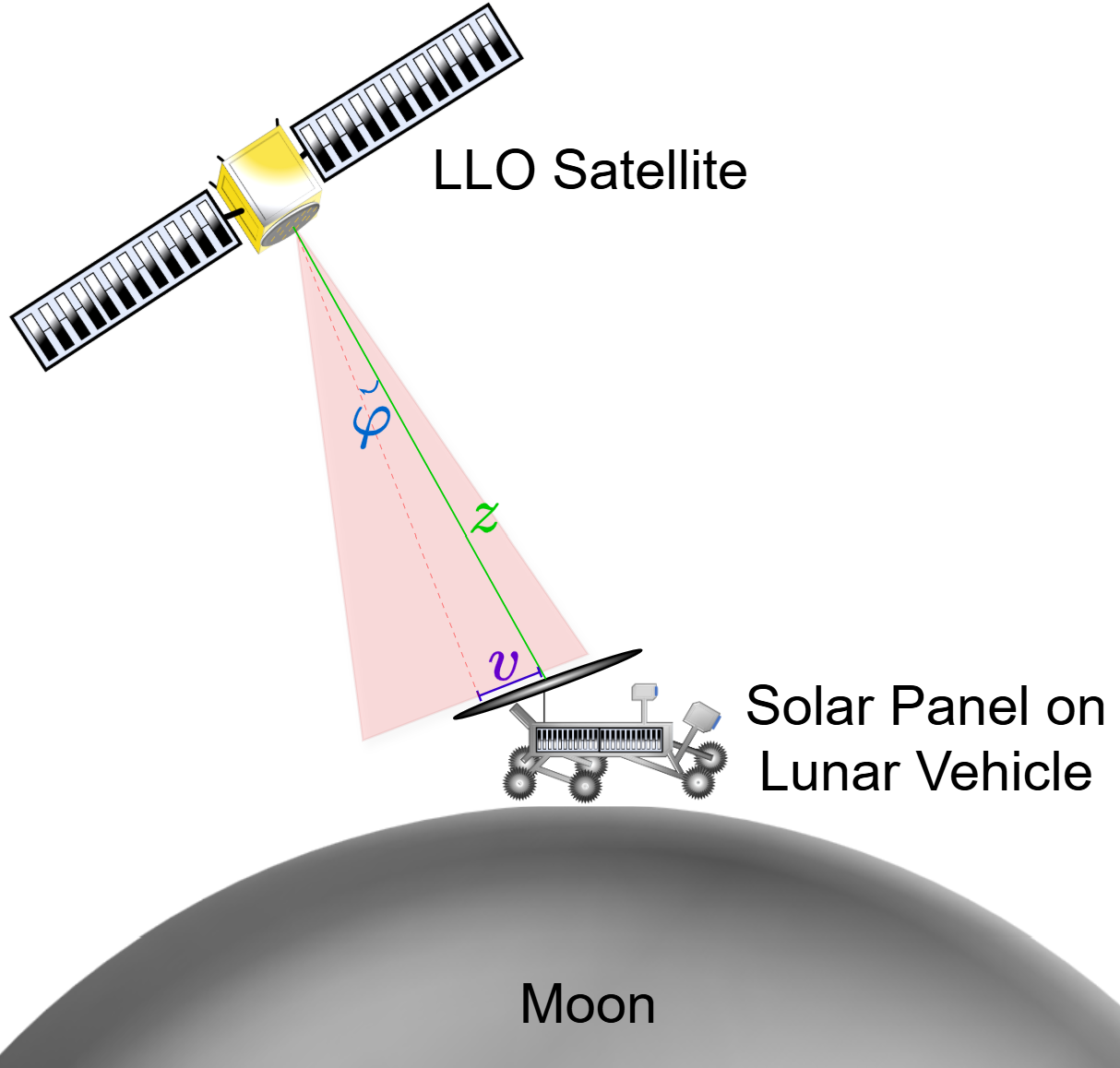}
\caption{Low lunar orbit satellite transmits a Gaussian beam to the circular solar panel mounted on the lunar vehicle with a distance $z$, radial error angle $\varphi$, and resulting spatial displacement error $v$.}
\label{FIG:2}
\vspace{-0.2cm}
\end{figure}

All these lunar orbits have benefits and drawbacks for lunar power beaming; however, a decision must be made with its reasons. Given that the LLO has no constraint on inclination \cite{orbittypes}, it can circle around the lunar poles, such as the promising Lunar South Pole (LSP) \cite{donmez2025hybridfsorflunar} as shown in Fig. \ref{FIG:2}. Moreover, the shorter the line-of-sight (LoS) distance, the smaller the beam footprint diameter (beam spot size), and hence the greater the collected power, given the limited receiver size. Although the LLO is less stable than NRO, which covers the poles, it offers the shortest lunar altitude and period, along with the flexibility of inclination; thus, it can be considered a better option for WPT on the Moon.

\subsection{Related Works}
\label{relatedworks}

The literature includes numerous works on FSO-based power transfer in Cislunar scenarios. In \cite{Txaperturedia}, the authors conduct a feasibility analysis of laser beaming from EMLP-1 (up to 59,000 km) and EMLP-2 (up to 66,000 km) to a manned rover on the Moon, targeting 30 kW of harvested power.
In addition, a deterministic misalignment error of 0.025 $\mu$rad is considered. According to \cite{Nasa2000km3sat}, three identical satellites, equally spaced in circular orbits 2000 km above the lunar surface, beam laser power to a lunar habitat and a rover, enabling the harvesting of 1 MW and 75 kW, respectively. In \cite{kerslake2008lunar}, the author investigates the overall wireless power transfer (WPT) efficiencies for surface-to-surface transmission on the Moon using copper cable, radiofrequency (RF), and FSO technologies. The objective is to transmit 10 kW of power over up to 1 km with a pointing error of 0.0286$^\circ$, and the results show that FSO outperforms RF at 5.8 GHz. In \cite{donmez2024continuous}, the authors examine a continuous power transmission to the Lunar Far Side from the halo orbits centred at EMLP-2, considering stochastic misalignment errors. In \cite{lopez2023lunar}, the authors investigate lunar constellations consisting of up to ten satellites across multiple orbital configurations, including a 600 km altitude lunar circular orbit, a lunar frozen orbit defined by specific Keplerian parameters for Shackleton Crater, and a DRO for the lunar equator. The study also analyzes battery usage. In \cite{landis2024power}, the study proposes a feasibility analysis of power beaming from three lunar orbiters to small landers situated between the equator and the pole. An altitude of 800 km yields transmission distances of up to 1600 km, with a long 35-minute visibility window. This study considers intermittent battery charging. 
According to \cite{donmez2025hybridfsorflunar}, the authors investigate a multi-hop hybrid wireless power transfer (WPT) system incorporating random misalignment errors. The harvested power at these destinations is modelled using stochastic methods to account for random pointing errors in the initial hop, which constitute a major limitation of FSO transmission. Various orbital scenarios based on LLO satellite deployment are investigated to establish a benchmark scenario in which continuous lunar power beaming is achieved for a full revolution, around 27.3 days, in \cite{donmez2026multi}. 

Furthermore, random misalignment errors are modelled frequently with Gaussian and Uniform distributions in space links \cite{Pointing1_casasco2013pointing, Pointing2_ott2011esa, Pointing3_hirth2017peet, Pointing4_casasco2014pointing, pointingerror}.  

The literature includes studies on wireless power transmission in space, deterministic or probabilistic pointing error models, and battery state-of-charge models. Nevertheless, our research explores a niche area where randomness affects the battery state-of-charge, accumulated battery capacity, accumulated battery energy, and end-to-end efficiency in realistic low lunar orbit. 

\subsection{Contributions}
\label{contributions}

In this work, we establish a laser-based wireless power transmission system in which a low-lunar orbiter beams a Gaussian beam to a circular solar panel mounted on the lunar vehicle at the Lunar South Pole, as shown in Fig. \ref{FIG:2}.

In our scenario, various levels of random transmitter pointing error arising from satellite wobbling are considered. Moreover, realistic time-varying distances and the corresponding beaming time $T_b$ with a horizon-mask constraint are computed. The tracking circular solar panel on the lunar vehicle is preferred to eliminate the diminishing effects of the angle of incidence \cite{donmez2026multi}. Furthermore, a space battery with realistic specifications exhibited in Table \ref{tab:SpaceBatteries} is preferred.

Our work realistically demonstrates how efficient lunar power beaming is under various realistic disturbances and how it affects the key battery metrics in pure fast-charging and slow-charging modes. We present statistical models of key metrics across multiple pointing-error levels, as well as scenarios with only fast or slow charging, to inform investment decisions. Each PDF provides a mathematical model that enables prediction and helps us understand the randomness of the corresponding variable. Note that, to make fair comparisons, a common set of bin edges is used for all three pointing error severities for each variable; this enables us to compare bin probability curves for state-of-charge, for instance.  Furthermore, the orbiter pass is repeated using the Monte Carlo simulation count shown in Table \ref{table:WPTSimParameters}, and we assume that a single randomly generated radial error is used for each orbiter revolution throughout the charging duration. Hence, the distributions are constructed across the passes.

Our contributions are as follows:
 \begin{itemize}

\item We define various pointing error bounds that represent severity. We then use the total width of these upper bounds to model zero-mean, Gaussian-distributed azimuth and elevation angles, which are the components of the Rayleigh-distributed radial error angle.

\item The stochastic distribution of harvested power for the lunar vehicle is determined in a realistic lunar environment established using STK software. Its random behaviour for each pointing-error severity is presented. The higher $\varphi_{c,max}$ causes larger variance and hence a larger standard deviation of the displacement error $v$. As a result, $P_h$ can reach up to 631 W with various bin probabilities, with mean values of 629.46, 589.02, and 490.60 W for high, medium, and low pointing-error levels, respectively.

\item As we extend our probabilistic analyses to the battery connected to the lunar vehicle, we derive the PDFs of the state of charge $SoC$, accumulated battery capacity $A_C$, accumulated battery energy $A_E$, and overall efficiency $\xi$. We then verify our derivations with Monte Carlo simulations.  

\item We quantify the differences between the battery charging modes; fast charging mode exhibits linearly increasing SoC behaviour, whereas the slow charging mode exhibits exponential decay.For instance, it is found that the mean of $A_{E,sc}$ is 13.34 Wh, which is around one-fifth of the mean $A_{E,fc}$ for $\varphi_{c,max}=5 \mu$rad.

\item In our Cislunar LLO WPT, we use STK's high-precision orbit propagator (HPOP), which computes ephemeris data by numerically integrating differential equations, to model realistic orbiter motion. The gravitational force models of third bodies (e.g., Earth), solar radiation pressure (SRP), and Moon radiation pressure (e.g., albedo) are considered. In addition, we apply the non-uniform lunar gravity GL0660B \cite{konopliv2013jpl} to surrounding objects, such as the orbiters in our scenarios.

\end{itemize}

\subsection{Organization}
\label{organization}
The remainder of this paper is organized as follows. Section II explains the Gaussian-profile laser beam, the random pointing-error model, and the resulting random battery state-of-charge model in detail. The PDF of harvested power and key battery metrics are derived and presented. In Section III, the Monte Carlo simulations are conducted to verify our mathematical derivations for pure fast and pure slow charging modes for comparison.  Finally, Section IV concludes the study.

%% file: Sections/SystemMod.tex
Shorter distances are desired for wireless laser transmission, as the spot size is directly proportional to distance. Therefore, LLO is a better option than other orbits because it is independent of inclination \cite{orbittypes}. The continuous-wave laser at 1064 nm uses a single-way path between the LLO satellite and the solar array mounted on the space vehicle at the Lunar South Pole.

As the satellite oscillates during the flyby, the Gaussian beam may experience pointing errors that displace the beam footprint center. Moreover, the incidence angle reduces the power collected by the solar panel.

\subsection{Laser Beam Modeling}

The Gaussian wave model is employed to characterize the spatial distribution of free-space laser beam irradiance at an axial distance $z$ between the satellite and the reflector or circular solar panel as follows \cite{majumdar2010free}: 

{\small
\begin{equation}
{I}(r,z)={{I_{0}}}\exp \left( \frac{-2{\,{r}^{2}}}{w{{(z)}^{2}}} \right)=\frac{2 P_{t,o}}{\pi w{{(z)}^{2}}}\exp \left( \frac{-2{\,{r}^{2}}}{w{{(z)}^{2}}} \right), 
\label{EQ:1} 
\end{equation}
}%

\noindent where $r$, $w{(z)}$, ${I}_{0}$, and $P_{t,o}$ are the radial distance from the center of the Gaussian beam footprint, Gaussian beam radius, the maximum irradiance at the beam center,  and optical transmit power, respectively.

The Gaussian beam radius $w{(z)}$ at distance $z$ is given as \cite{majumdar2010free}.

{
\begin{equation}
w(z)={{w}_{0}}\sqrt{1+{{\left( \frac{z\lambda }{\pi {{w}_{0}}^{2}} \right)}^{2}}}, 
\label{EQ:2} 
\end{equation}
}%

\noindent where ${w}_{0}$ and $\lambda$ are the beam waist and the operational wavelength, respectively.

The Gaussian beam waist ${w}_{0}$ and the beam divergence angle $\theta$ are described as follows \cite{OWCmatlab}:

\begin{align} 
& {{w}_{0}}=\frac{\lambda }{\pi \theta }, \\[10pt] 
& \theta\cong \lambda /{{D}_{t}}. 
\label{EQ:3and4}
\end{align}

\noindent The larger the laser telescope aperture diameter $D_t$, the narrower the beam divergence angle $\theta$.


The collected portion of the Gaussian beam transmitted by the satellite is calculated as follows \cite{farid2007outage}  

{
\begin{equation}
{{{P}_{c}}}=\int\limits_{A'}{\,I(r,z)\, dA}.
\label{EQ:5} 
\end{equation}
}%

The circular solar panel harvests the Gaussian beam, as shown in Fig. \ref{FIG:2}. Given the solar panel area of $A$, the fraction of the collected optical power is formulated by

{\small
\begin{equation}
 {{P}_{c}}=\int\limits_{\phi'}{\int\limits_{r'}{\frac{2P_{t,o}}{\pi w{(z)^{2}}}\exp \left( \frac{-2\,{{r}^{2}}}{w{(z)^{2}}} \right)\,r\,dr\,d\phi }}. 
 \label{EQ:6} 
 \end{equation}
 }%

\begin{figure}[t!]
\centering
\includegraphics[width=.85\columnwidth]{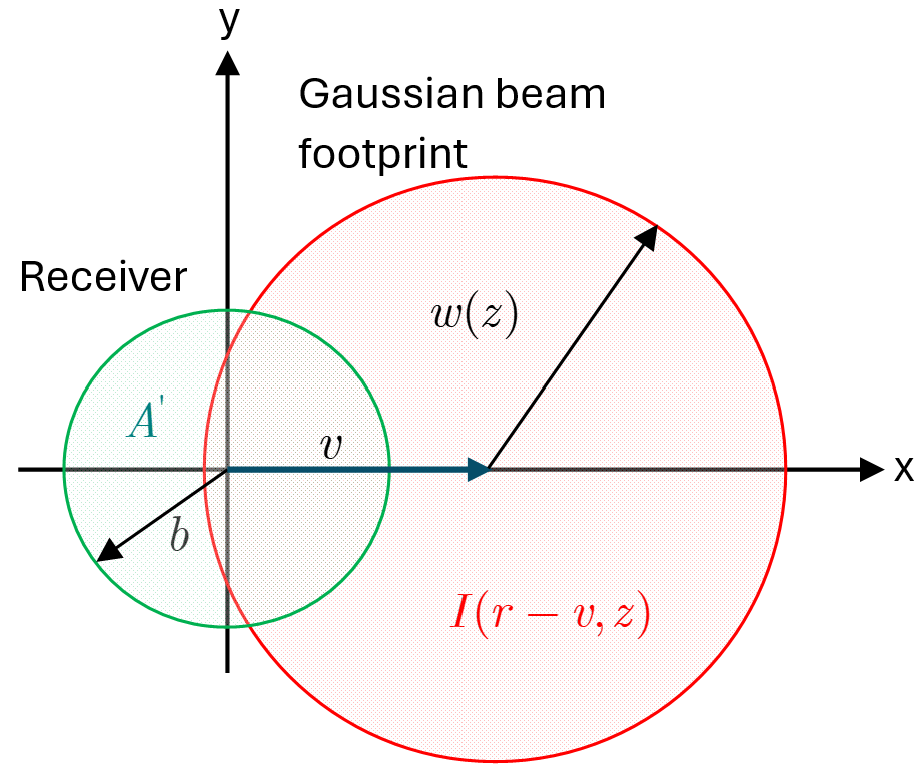}
\caption{Displacement $v$ from the pointing error shows the misalignment between the receiver centre (radius $b$) and the swayed irradiance $I(r-v,z)$ with beam radius $w(z)$ \cite{donmez2025hybridfsorflunar}.}
\label{FIG:3}
\vspace{-0.2cm}
\end{figure}
 
Fig. \ref{FIG:3} shows the spatial displacement error $v$ caused by pointing error at the satellite transmitter. This misalignment reduces the intersection area between the Gaussian beam footprint and the receiver, thereby degrading the collected optical power of energy-transfer signals; the greater the misalignment, the smaller the portion of the beam collected.

The received power with misalignment error can be rewritten as follows \cite{donmez2025hybridfsorflunar}: 

{\small
\begin{equation}
{{P}_{{c}}}=\int\limits_{-b}^{b}{\int\limits_{-\sqrt{{{b}^{2}}-{{x}^{2}}}}^{\sqrt{{{b}^{2}}-{{x}^{2}}}}{\frac{2{P_{t,o}}}{\pi w{{(z)}^{2}}}}}\exp \left( -2\,\frac{{{(x-v)}^{2}}+{{y}^{2}}}{w{{(z)}^{2}}} \right)\,\,dy\,dx.
\label{EQ:7}
\end{equation}
}

For a given $w(z)/b>6$ (i.e., normalized mean square error less than $10^{-3}$ \cite{farid2007outage}), an analytical approximation $P_a$ of the received power $P_c$ provides a faster solution from the integration in Eq. (\ref{EQ:7}). The analytical formula is given by \cite{farid2007outage}:

\begin{align}
\label{EQ:8}
& c_1=\frac{b\sqrt{\pi }}{{{w}({z})}\sqrt{2}}, \,\,\, {{C}_{0}}={{\left[ \text{erf}(c_1) \right]}^{2}}  \nonumber, \\[10pt] 
& {w(z_{eq})^{2}}={w(z)^{2}}\left[ \frac{\sqrt{\pi }\,}{2\,c_1\,}\times \frac{\text{erf}(c_1)}{\exp (-{{c_1}^{2}})} \right], \\[10pt]
& {{P}_{a}}\approx {P_{t,o}}\,{{C}_{0}}\exp \left( \frac{-2{{v}^{2}}}{w({z_{eq}})^{2}} \right). \nonumber
\end{align}
\noindent To find the collected optical power, we must multiply by the optical transmit power $P_{t,o}$, since the multiplicative factor on the right-hand side gives the fraction of transmitted optical power that is collected.

Finally, the harvested electrical power $P_h$ can be found with using the optical-to-electrical PCE $\eta_h$ as follows
\begin{equation}
 {{P}_{h}} = { \eta_h \times {{P}_{a}} }. 
\label{EQ:9}
\end{equation}

\begin{table}[!t]
  \sisetup{group-minimum-digits = 4}
  \centering
  \caption{WPT Simulation Parameters}
  \label{table:WPTSimParameters}
  \begin{tabular}{lllS[table-format=5]ll} 
    \toprule
    \toprule
\begin{tabular}[c]{@{}l@{}}Laser signal electrical \\ power\end{tabular} & 50 kW\\
\begin{tabular}[c]{@{}l@{}}Laser electro-optical \\ conversion efficiency \end{tabular} & 51\% \\
\begin{tabular}[c]{@{}l@{}}Satellite telescope aperture\\ diameter \end{tabular} & 0.1 m\\
\begin{tabular}[c]{@{}l@{}}Laser wavelength \end{tabular} & 1064 nm\\
\begin{tabular}[c]{@{}l@{}}Circular solar panel \\ diameter \end{tabular} & 1.4 m \\
\begin{tabular}[c]{@{}l@{}}Solar cell efficiency \end{tabular} & 50.8\% \\
\begin{tabular}[c]{@{}l@{}}Horizon mask\\\end{tabular} & 5$^\circ$ \\
\begin{tabular}[c]{@{}l@{}}Number of bins \end{tabular} & 1000 \\ 
\begin{tabular}[c]{@{}l@{}}Monte Carlo simulation \\ number\end{tabular} & 10$^7$ \\
   \bottomrule
    \bottomrule
\end{tabular}
\end{table} 

\begin{figure}[t!]
\centering
\includegraphics[width=.9\columnwidth]{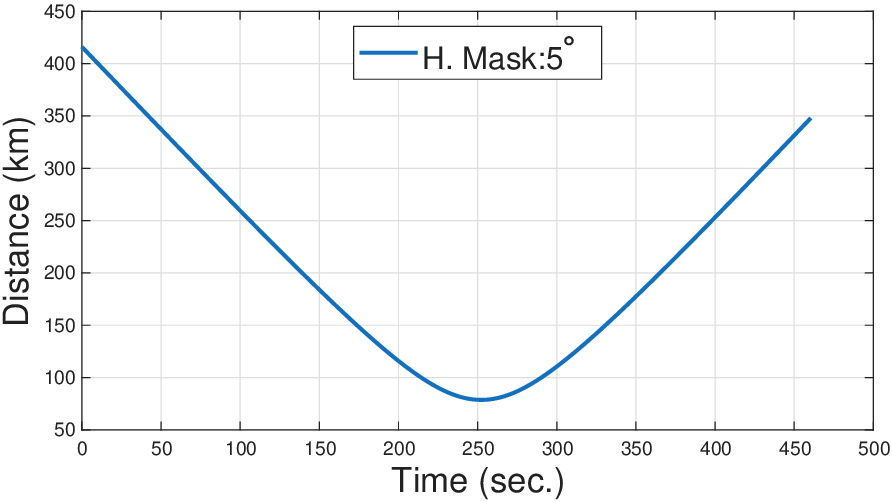}
 \caption{Time-varying satellite-to-lunar-vehicle distance.}
 \label{FIG:4}
 \end{figure}

Up to this point, we have expounded the Gaussian beam propagation equation and its corresponding analytical formula. In the following subsection, we model zero-mean, Gaussian-distributed azimuth and elevation error angles, as well as the resulting radial error angle and spatial displacement error, which follow the Rayleigh distribution for three misalignment-error levels. It is crucial to analyze the random harvested powers at the maximum distance during the flyby. Because the spatial displacement error is maximized along with the distance, our work considers a worst-case scenario for given pointing-error limits. In other words, our work demonstrates the lower limits of system performance, which improves as $z$ shortens and $v$ reduces. Fig. \ref{FIG:4} shows the time-varying path length. The farthest and nearest target distances are computed as approximately 416 km and 78.7 km, respectively.  
For the given simulation setting, this maximum distance yields a beam footprint radius $w(z)$ of 4.43 m. Therefore, $w(z)/b=$ 6.32, which is larger than 6 and within the approximation validity regime \cite{farid2007outage}.

\subsection{Stochastic Pointing Error Modeling }
\label{Sec:sysmod:PointingError}

Pointing errors arise solely from satellite wobbling, as the receiver on the Moon remains stationary in the Moon-centred, Moon-fixed reference frame. The horizontal and vertical components of the radial error angle are the azimuth error angle $\varphi_{az}$ and the elevation error angle $\varphi_{el}$, respectively. Given the upper limits, these error angles can be modelled stochastically. 

The pointing-error upper limits $\varphi_{c,max}$ for the components are set to 0.5 $\mu$rad, 2.5 $\mu$rad, and 5 $\mu$rad, corresponding to low (mild), medium (moderate), and high (severe) conditions, respectively. Thus, the pointing errors on each axis are uniformly distributed in $\pm$0.5, $\pm$2.5, and $\pm$5 $\mu$rad with total widths of 1 $\mu$rad, 5 $\mu$rad, and 10 $\mu$rad, respectively, while modelling the zero-mean Gaussian distribution variance. Given that $\varphi_c\in\{\varphi_{az},\varphi_{el}\}$, the equivalent uniform bound is utilized when finding the variance of the Gaussian-distributed azimuth and elevation error angles as follows \cite{leongarcia2008}
\begin{equation}
\sigma _{\varphi_c}^{2}=\frac{(2\varphi _{c,\max })^{2}}{12}.
\label{EQ:10}
\end{equation}

\subsubsection{PDF of Radial Error}


The zero-mean Gaussian-distributed azimuth error angle $\varphi_{az}$ and elevation error angle $\varphi_{el}$ are the components of the radial error angle ${\varphi}=\sqrt{\varphi _{az}^{2}+\varphi _{el}^{2}}$  where ${{\varphi }_{az}},{{\varphi }_{el}}\sim N(0,\sigma _{\varphi }^{2})$. The PDF of the radial error angle follows a Rayleigh distribution and is given by \cite{Shlomi_optimization_2004}
 
\begin{equation}
{{f}_{\varphi }}(\varphi )=\frac{\varphi }{\sigma _{\varphi }^{2}}\exp \left( -\frac{{{\varphi }^{2}}}{2\sigma _{\varphi }^{2}} \right),\,\,\,\,\,\,\,\varphi \ge 0
\label{EQ:13}
\end{equation}



\subsubsection{PDF of Spatial Displacement Error}
The spatial displacement $v$ is simply the multiplication of the LoS distance $z$ and the radial error angle $\varphi$ in radians, owing to the small-angle approximation. The PDF of the spatial displacement $v$ can be found with ${{\sigma }_{v}}=z\,{{\sigma }_{\varphi }}$ as follows
\begin{equation}
{{f}_{v}}(v)=\frac{v}{\sigma _{v}^{2}}\exp \left( -\frac{{{v}^{2}}}{2\sigma _{v}^{2}} \right),\,\,\,\,\,\,\,v\ge 0
\label{EQ:14}
\end{equation}

\subsubsection{PDF of Harvested Power}
From Eq. (\ref{EQ:8}) and (\ref{EQ:9}) the inverse function, to be able to express $v$ in terms of $P_h$, is computed as follows
\begin{align}
  & \frac{{{P}_{h}}}{{{\eta }_{h}}{{P}_{t}}{{C}_{0}}}=\exp \left( \frac{-2}{w{{({{z}_{eq}})}^{2}}}\cdot {{v}^{2}} \right), \nonumber \\ 
 &  \nonumber\\ 
 & v=\frac{w({{z}_{eq}})}{\sqrt{2}}\sqrt{\ln \left( \frac{{{\eta }_{h}}{{P}_{t}}{{C}_{0}}}{{{P}_{h}}} \right)}. \nonumber 
\label{EQ:15}
\end{align}

The Jacobian factor is found by
\begin{align}
  & \frac{d{{P}_{h}}}{dv}={{\eta}_{h}}\,{P_{t,o}}\,{{C}_{0}}\left[ \exp \left( \frac{-2{{v}^{2}}}{w{{({{z}_{eq}})}^{2}}} \right)\cdot \frac{-4v}{w{{({{z}_{eq}})}^{2}}} \right], \nonumber \\ 
  \\
 & =\frac{-4v}{w{{({{z}_{eq}})}^{2}}}\cdot {{P}_{h}}, \nonumber
\label{EQ:16}
\end{align}

\noindent then the reciprocal of $dP_h/dv$ is found. 
\begin{equation}
\left| \frac{dv}{d{{P}_{h}}} \right|=\left| {{\left( \frac{d{{P}_{h}}}{dv} \right)}^{-1}} \right|=\frac{w{{({{z}_{eq}})}^{2}}}{4v\,{{P}_{h}}}.
\label{EQ:17}
\end{equation}

Applying the transformation of variables theorem \cite{leongarcia2008} to obtain the PDF of the electrical harvested power $P_h$ as follows
\begin{align}
  & {{f}_{{{P}_{h}}}}\left( {{p}_{h}} \right)={{f}_{v}}\left( v\left( {{p}_{h}} \right) \right)\,\,\left| \frac{dv}{d{{p}_{h}}} \right|, \nonumber \\ 
 &  \nonumber \\ 
 & ={{f}_{v}}\left( \frac{w({{z}_{eq}})}{\sqrt{2}}\sqrt{\ln \left( \frac{{{\eta }_{h}}{{P}_{t,o}}{{C}_{0}}}{{{p}_{h}}} \right)} \right)\times \frac{w{{({{z}_{eq}})}^{2}}}{4v\,{{p}_{h}}}, \\ 
 &  \nonumber \\ 
 & =\left[ \frac{\sqrt{2}\alpha }{{{\sigma }_{v}}}\sqrt{\ln \left( \frac{{{\eta }_{h}}{{P}_{t,o}}{{C}_{0}}}{{{p}_{h}}} \right)}{{\left( \frac{{{\eta }_{h}}{{P}_{t,o}}{{C}_{0}}}{{{p}_{h}}} \right)}^{-{{\alpha }^{2}}}} \right]\times  \nonumber \\ 
 & \frac{w{{({{z}_{eq}})}^{2}}}{4v\,{{P}_{h}}}. \nonumber
\label{EQ:18}
\end{align}

\noindent where $\alpha ={w({{z}_{eq}})}/{(2{{\sigma }_{v}})}\;$.

After cancellations and simplifications, it can be written as
\begin{equation}
{{f}_{{{P}_{h}}}}\left( {{p}_{h}} \right)=\frac{{{\alpha }^{2}}}{{{\left( {{\eta }_{h}}{{P}_{t,o}}{{C}_{0}} \right)}^{{{\alpha }^{2}}}}}p_{h}^{{{\alpha }^{2}}-1}.
\label{EQ:19}
\end{equation}
\\
\noindent where $0\le {{p}_{h}}\le {{\eta }_{h}}{{P}_{t,o}}{{C}_{0}}$.

Up to this point, we conducted stochastic analyses of power transmission between the LLO satellite and the solar panel mounted on the lunar vehicle. From now on, we focus on the stochastic storage model because the harvested power fed to the battery inside the rover behaves randomly. As per \cite{AirbusBattery}, a similar battery can weigh around 3 kg.


\subsection{Stochastic Battery SoC Modeling}
\label{Sec:sysmod:Battery}

The unitless battery state of charge (SoC) is the ratio of accumulated capacity $A_C$ (ampere-hours) to the maximum battery capacity $C_b$. Time-varying SoC increases linearly with a constant charging current $I_b$ (e.g., $1.5 C$) until it reaches $SoC_{FS}$; then the battery saturates and charging continues inefficiently (e.g., $C/5$) while maintaining a constant voltage in this slow-charging phase. The conventional C-rate notation exhibited in Table \ref{tab:SpaceBatteries} is the numerical magnitude of $C_{b}$ (Ah) \cite{AirbusBattery}; hence, the minimum charging current becomes 30 A for slow-charging mode. Note that, to establish a pure slow-charging battery mode, $SoC_{FS}$ can be limited to the initial SoC $SoC_{0}$, which is defined by the battery depth-of-discharge (DoD) (i.e., $SoC_{0}=1-DoD$); thus, the fast-charging state can be omitted as only slow-charging mode is applicable.

Fast Charging Phase: In this phase, the maximum $I_b$ shown in Table \ref{tab:SpaceBatteries} for the fast-charging state, $I_{b,fc}$, is maintained, and the SoC increases linearly with time. When the SoC converges to the Fast-to-Slow charging mode transition SoC $SoC_{FS}$.

The SoC(t) in the fast charging phase is evaluated using the Coulomb counting method \cite{bergveld2002battery} as follows
\begin{equation}
So{{C}_{fc}}=So{{C}_{0}}+\left( \frac{{{I}_{b,fc }}}{{{C}_{b}}}\cdot t \right). 
\label{EQ:24}
\end{equation}

The required amount of time to reach from $SoC_0$  to $SoC_{FS}$ can be computed by
\begin{equation}
 {{T}_{fc}}=\frac{{{C}_{b }}}{{{I}_{b,fc}}}\times (So{{C}_{FS}}-So{{C}_{0}})\,\,.  
\label{EQ:25}
\end{equation}

Slow Charging Phase: The linearly increasing charging behaviour transitions to an increasing, concave-down curve, which is inefficient for battery recharging.  This is because the charging current decays exponentially as the battery saturates.

The SoC in the slow charging phase is computed using the Coulomb counting method \cite{bergveld2002battery} as follows

\begin{align}
  & So{{C}_{sc}}=So{{C}_{FS}}+\left( \frac{{{I}_{b,sc}}\cdot \tau }{{{C}_{b}}} \right)\times  \nonumber \\ 
 & \left( 1-\exp \left( -\frac{t-{{T}_{fc}}}{\tau } \right) \right) 
\label{EQ:26and27}
\end{align}


\noindent The variable $\tau$ is the RC time constant and is considered to be 1 hour.


The piecewise mathematical model of SoC(t) is given as
\begin{equation}
SoC=
 \begin{cases}
  So{{C}_{fc}} & \text{if }  SoC\le {{SoC_{FS}}}  
  \\
  So{{C}_{sc}} & \text{if } SoC>{{SoC_{FS}}}  
\end{cases}.
\label{EQ:28}
\end{equation}

\begin{table}[t!]
\centering
\caption{Space Battery Parameters Used in Simulations \cite{AirbusBattery}}
\label{tab:SpaceBatteries}
\resizebox{\columnwidth}{!}{%
\begin{tabular}{|c|c|c|c|c|}
\hline
  \begin{tabular}[c]{@{}c@{}}Nominal \\ Voltage \\ (V)\end{tabular} &
  \begin{tabular}[c]{@{}c@{}}Capacity \\ (Ah)\end{tabular} &
  \begin{tabular}[c]{@{}c@{}}Depth of \\ Discharge \\ (\%)\end{tabular} & 
  \begin{tabular}[c]{@{}c@{}}Max. Charging  \\ Current (fc) \\(A)\end{tabular} & 
  \begin{tabular}[c]{@{}c@{}}Max. Charging  \\ Current (sc)\\(A)\end{tabular}
  \\ \hline
 3.7 & 150  & 80 & 1.5C & C/5 
 \\ \hline
\end{tabular}%
}
\end{table}

\subsubsection{PDF of State of Charge}
The PDF of SoC serves as a mathematical model that facilitates prediction and helps us understand its randomness. Since the model of SoC is piecewise, there will be two separate PDF analyses for the conditions $SoC \le SoC_{FS}$ and $SoC>SoC_{FS}$. The PDF of the $SoC$ is transformed from the PDF of the harvested power $P_h$.


Condition 1: $SoC \le SoC_{FS}$
The relationship between the $SoC$ and $P_h$ is presented to find the inverse function and then the first-order derivative. 
\begin{align}
& SoC=So{{C}_{0}}+\frac{({{P}_{h}}/{{V}_{b}}){{T}_{b}}}{{{C}_{b}}} \nonumber ,\\
 &  \\ 
  & SoC=So{{C}_{0}}+\beta_{fc} \cdot {{P}_{h}},  \nonumber\\ 
 &  \nonumber \\ 
 & \beta_{fc} =\frac{{{T}_{b}}}{{{V}_{b}}\,{{C}_{b}}}. 
\label{EQ:29and30}
\end{align}
\noindent where $V_b$ is the nominal voltage of 3.7 V. 

The multiplying factor of $P_h$ is defined as $\beta_{fc}$, which is specific to the fast-charging mode.

Condition 2: $SoC>SoC_{FS}$
\begin{align}
  & SoC=SoC_{FS}+\frac{{{P}_{h}}\tau}{{{V}_{b}}C_b}\times\left( 1-e^{-({T}_{b}/\tau)} \right)   \\ \nonumber
  &  \\ 
  & SoC=So{{C}_{FS}}+\beta_{sc} \cdot {{P}_{h}},  \nonumber\\ 
 &  \nonumber
\label{EQ:31}
\end{align}
The multiplier of $P_h$ is defined as $\beta_{sc}$, which is peculiar to the slow-charging mode.
\begin{equation}
{{\beta }_{sc}}=\frac{\tau }{{{V}_{b}}{{C}_{b}}}\times \left( 1-\exp \left( -\frac{ {{T}_{b}} }{\tau } \right) \right)
\label{EQ:32}
\end{equation}

The inverse function and then the Jacobian are computed as follows 

 \begin{figure}[!t]
 \centering
\subfloat[]{
 	\label{subfig:5a}
 	\includegraphics[clip, width=0.95\columnwidth]{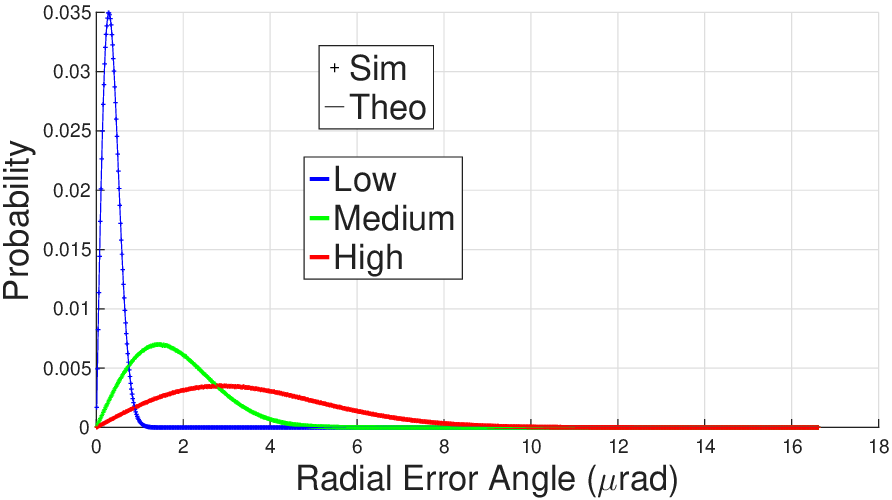}
 	 }
     \\
\subfloat[]{
 	\label{subfig:5b}
 	\includegraphics[clip, width=0.95\columnwidth]{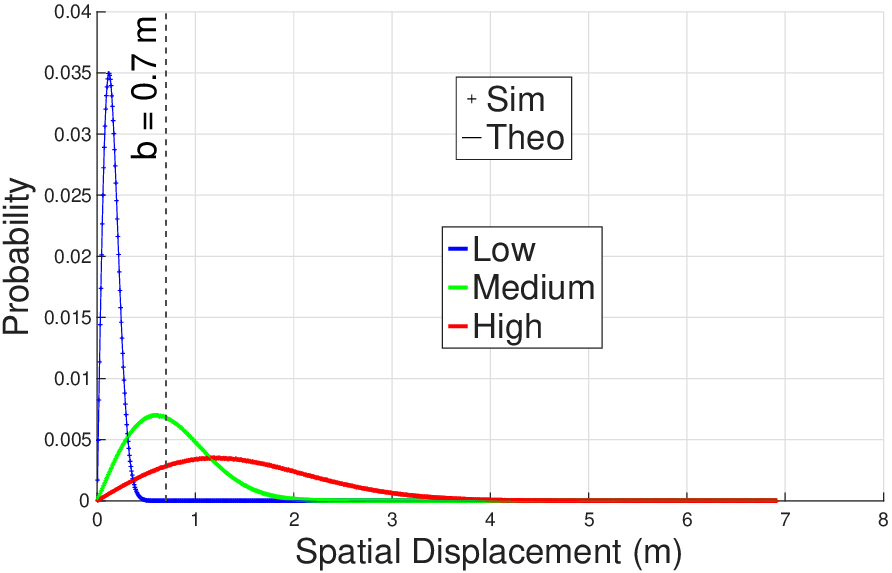}
 	 }
     \\
\subfloat[]{
 	\label{subfig:5c}
 	\includegraphics[clip, width=0.95\columnwidth]{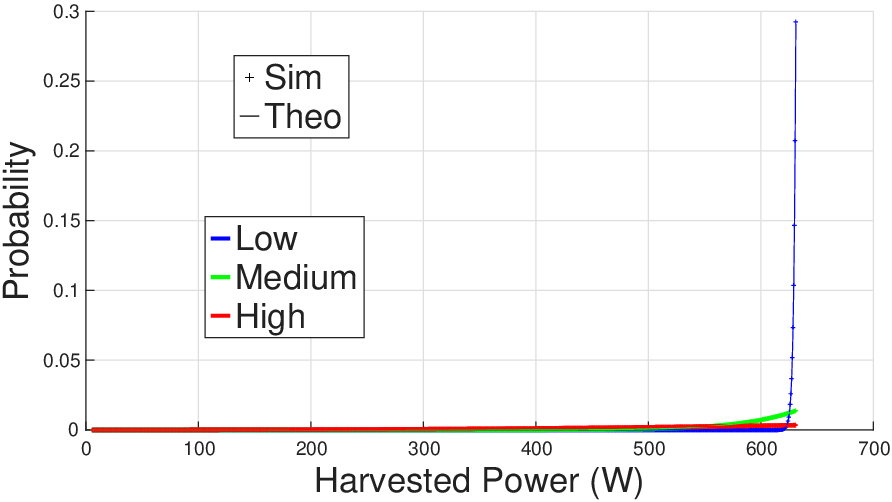}
 	 }
 \caption{Monte Carlo simulation versus theoretical PDF for low to high pointing errors: (a) radial error angle distribution, (b) spatial displacement error distribution, (c) harvested power distribution.}
 \label{FIG:5}
 \end{figure} 

\begin{equation}
{{f}_{SoC}}(s)={{f}_{{{P}_{h}}}}\left( {{P}_{h}}(s) \right)\times \left| \frac{d{{P}_{h}}}{ds} \right|
\label{EQ:33}
\end{equation}

\begin{align}
  & {{P}_{h}}(s)=\frac{s-So{{C}_{0}}}{\beta } \\ 
 & \left| \frac{d{{P}_{h}}}{ds} \right|=\frac{1}{\beta } 
 \label{EQ:34and35}
 \end{align}

\begin{equation}
{{f}_{{{P}_{h}}}}\left( {{P}_{h}}(s) \right)=\frac{{{\alpha }^{2}}}{{{\left( {{\eta }_{h}}{P_{t,o}}{{C}_{0}} \right)}^{{{\alpha }^{2}}}}}\cdot \frac{{{\left( s-So{{C}_{0}} \right)}^{{{\alpha }^{2}}-1}}}{{{\beta }^{{{\alpha }^{2}}}}{{\beta }^{-1}}}
\label{EQ:36}
\end{equation}
\\
\noindent where $\alpha ={w({{z}_{eq}})}/{(2{{\sigma }_{v}})}\;$,  $\beta\in\{\beta_{fc},\beta_{sc}\}$, and the PDF argument $s$ is the exact realization of $SoC$ random variable.

\begin{align}
  & {{f}_{SoC}}(s)=\left( \frac{{{\alpha }^{2}}}{{{\left( {{\eta }_{h}}{P_{t,o}}{{C}_{0}} \right)}^{{{\alpha }^{2}}}}}\cdot \frac{{{\left( s-So{{C}_{0}} \right)}^{{{\alpha }^{2}}-1}}}{{{\beta }^{{{\alpha }^{2}}}}} \right) \nonumber \\ \nonumber
 &  \\ 
 & =\frac{{{\alpha }^{2}}}{{{\left( {{\eta }_{h}}{P_{t,o}}{{C}_{0}}\beta  \right)}^{{{\alpha }^{2}}}}}\cdot {{\left( s-So{{C}_{0}} \right)}^{{{\alpha }^{2}}-1}}
\label{EQ:37}
\end{align}
 
Since the function ${{f}_{SoC}}(s)$ includes a $ \beta$ variable, there are two separate PDFs for fast and slow charging conditions, as follows.

\begin{equation}
 {f}_{SoC}(s)=\begin{cases}
  \frac{{{\alpha }^{2}}{{\left( s-So{{C}_{0}} \right)}^{{{\alpha }^{2}}-1}}}{{{\left( {{\eta }_{h}}{P_{t,o}}{{C}_{0}} \right)}^{{{\alpha }^{2}}}}}\cdot \beta _{fc}^{-{{\alpha }^{2}}},\,\,\,\,\,\,\,\,\,SoC\le SOC_{FS} 
  \\ 
 \frac{{{\alpha }^{2}}{{\left( s-So{{C}_{0}} \right)}^{{{\alpha }^{2}}-1}}}{{{\left( {{\eta }_{h}}{P_{t,o}}{{C}_{0}} \right)}^{{{\alpha }^{2}}}}}\cdot \beta _{sc}^{-{{\alpha }^{2}}},\,\,\,\,\,\,\,\,\,SoC>SOC_{FS} \\ 
\end{cases} 
\label{EQ:38}
\end{equation}

\noindent where
\begin{align}
 & So{{C}_{\min }}\le s \le So{{C}_{\max }}  \nonumber \\ 
 &  \nonumber \\ 
 & So{{C}_{\min }}=So{{C}_{0}}  \\ \nonumber
 &  \\ \nonumber
 & So{{C}_{\max }}=So{{C}_{0}}+{{\eta }_{h}}{P_{t,o}}{{C}_{0}}\beta  \nonumber
\end{align}
 
\subsubsection{PDF of Accumulated Battery Capacity}
The accumulated battery capacity is an essential metric that enables an evaluation between other batteries with the same $V_b$ since a high SoC information must be interpreted together with the battery features (e.g., smaller/larger capacity). The inverse function, to be able to express the $SoC$ in terms of $A_C$, is computed as follows
\begin{align}
  & {{A}_{C}}=\left( SoC-So{{C}_{0}} \right)\times {{C}_{b }}, \\ 
 &  \nonumber \\ 
 & SoC=So{{C}_{0}}+\frac{{{A}_{C}}}{{{C}_{b }}}, \\ 
 &  \nonumber \\ 
 & \left| \frac{dSoC}{d{{A}_{C}}} \right|=\frac{1}{{{C}_{b }}}  
\end{align}

The PDF of accumulated capacity, which is a function of the PDF of SoC, is presented as follows
\begin{align}
  & {{f}_{{{A}_{C}}}}({{a}_{c}})=\frac{1}{{{C}_{b }}}\times {{f}_{SoC}}\left( So{{C}_{0}}+\frac{{{a}_{c}}}{{{C}_{b }}} \right), \nonumber \\ 
 &  \\ 
 & {{f}_{{{A}_{C}}}}({{a}_{c}})=\frac{1}{{{C}_{b }}}\times \frac{{{\alpha }^{2}}{{\left( {{{a}_{c}}}/{{{C}_{b }}}\; \right)}^{{{\alpha }^{2}}-1}}}{{{\left( {{\eta }_{h}}\,{P_{t,o}}\,{{C}_{0}} \right)}^{{{\alpha }^{2}}}}}\cdot {{\beta }^{-{{\alpha }^{2}}}}  \nonumber
\end{align}

\noindent where
\begin{align}
  & {{A}_{C,}}_{\min }\le {a}_{c} \le {{A}_{C,}}_{\max } \nonumber\\ 
 &  \nonumber\\ 
 & {{A}_{C,}}_{\min }=0 \\ 
 &  \nonumber\\ 
 & {{A}_{C,}}_{\max }={{C}_{b }}\cdot \left( {{\eta }_{h}}\,{P_{t,o}}\,{{C}_{0}}\,\beta  \right)  \nonumber
\end{align}

The PDF of accumulated battery capacity provides a mathematical model for predicting and analyzing its inherent randomness.

 \begin{figure}[!t]
 \centering
\subfloat[]{
 	\label{subfig:6a}
 	\includegraphics[clip, width=0.95\columnwidth]{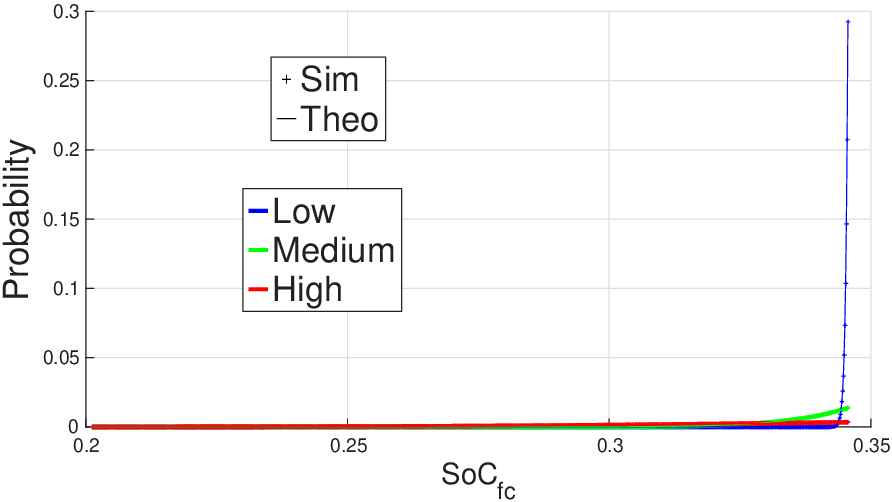}
 	 }
\\
\subfloat[]{
 	\label{subfig:6b}
 	\includegraphics[clip, width=0.95\columnwidth]{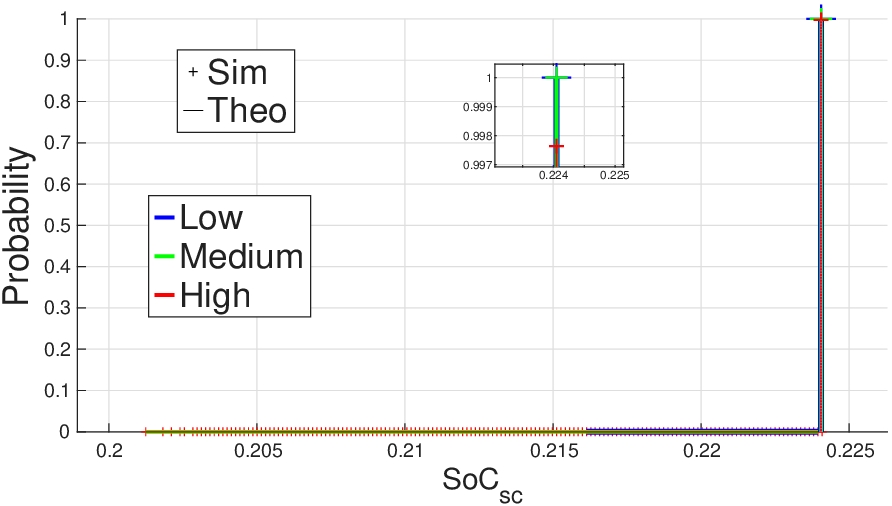}
 	 }
 \caption{Monte Carlo simulation versus theoretical PDF for low to high pointing errors: (a) fast charging $SoC$, (b) slow charging $SoC$.}
 \label{FIG:6}
 \end{figure}
 
\subsubsection{PDF of Accumulated Battery Energy}
The accumulated battery energy is a vital performance metric as it allows for a comparison regardless of the battery features such as capacity and voltage. The inverse function, to be able to express the $A_C$ in terms of $A_E$, is computed as follows
\begin{align}
  & {{A}_{E}}={{A}_{C}}\times V, \\ 
 &  \nonumber\\ 
 & {{A}_{C}}=\frac{{{A}_{E}}}{V}, \\ 
 &  \nonumber\\ 
 & \left| \frac{d{{A}_{C}}}{d{{A}_{E}}} \right|=\frac{1}{V}  
\end{align}

The change of variables formula is used to obtain the new PDF of the accumulated battery energy as follows 
\begin{align}
  & {{f}_{{{A}_{E}}}}({{a}_{e}})=\frac{1}{V}\times {{f}_{{{A}_{C}}}}\left( \frac{{{a}_{e}}}{V} \right), \nonumber\\ 
 &  \\ 
 & {{f}_{{{A}_{E}}}}({{a}_{e}})=\frac{1}{{{C}_{b }}V}\times \frac{{{\alpha }^{2}}{{\left( {{{a}_{e}}}/{\left( {{C}_{b }}V \right)}\; \right)}^{{{\alpha }^{2}}-1}}}{{{\left( {{\eta }_{h}}\,{P_{t,o}}\,{{C}_{0}} \right)}^{{{\alpha }^{2}}}}}\cdot {{\beta }^{-{{\alpha }^{2}}}} \nonumber 
\end{align}

\noindent where
\begin{align}
  & {{A}_{E,}}_{\min }\le {{a}_{e}} \le {{A}_{E,}}_{\max } \nonumber\\ 
 &  \nonumber\\ 
 & {{A}_{E,}}_{\min }=0 \\ 
 &  \nonumber\\ 
 & {{A}_{E,}}_{\max }={{C}_{b }}\,V\cdot \left( {{\eta }_{h}}\,{P_{t,o}}\,{{C}_{0}}\,\beta  \right) \nonumber 
\end{align}

The probability density function (PDF) of the accumulated battery energy serves as a mathematical model that facilitates prediction and clarifies its inherent variability.

\subsubsection{PDF of End-to-End Efficiency}
The end-to-end system efficiency $\xi$ is the ratio of the accumulated energy $A_E$ in the lunar battery to the transmitted energy, defined as the transmitted electrical power $P_{t,e}$ summed over the beaming time interval $T_b$.

The inverse function, to be able to express $\xi$ in terms of $A_E$, is computed as follows
\begin{align}
  & {{\xi }}=\frac{{{A}_{E}}}{\left( {{P}_{t,e}}\,{{T}_{b}} \right)}, \\ 
 &  \nonumber\\ 
 & {{A}_{e}}={{\xi }}\left( {{P}_{t,e}}\,{{T}_{b}} \right), \\ 
 &  \nonumber\\ 
 & \left| \frac{d{{A}_{E}}}{d{{\xi }}} \right|=\left( {{P}_{t,e}}\,{{T}_{b}} \right).  
\end{align}

The PDF of end-to-end efficiency, which is a function of the PDF of accumulated energy, is presented as follows
\begin{align}
  & {{f}_{{{\xi }}}}(\xi )={{P}_{t,e}}\,{{T}_{b}}\times {{f}_{{{A}_{E}}}}\left( {{P}_{t,e}}\,{{T}_{b}}\,\xi  \right), \nonumber \\ 
 &  \nonumber\\ 
 & {{f}_{{{\xi }}}}(\xi )=\frac{{{P}_{t,e}}\,{{T}_{b}}}{\,{{C}_{b }}\,V}\times  \\ 
 & \frac{{{\alpha }^{2}}{{\left( {{{P}_{t,e}}\,{{T}_{b}}\,\xi }/{\left( \,{{C}_{b }}\,V \right)}\; \right)}^{{{\alpha }^{2}}-1}}}{{{\left( {{\eta }_{h}}\,{P_{t,o}}\,{{C}_{0}} \right)}^{{{\alpha }^{2}}}}}\cdot {{\beta }^{-{{\alpha }^{2}}}} \nonumber 
\end{align}

\noindent where
\begin{align}
  & {{\xi }}_{\min }\le  \xi  \le {{\xi }}_{\max }, \nonumber\\ 
 &  \nonumber\\ 
 & {{\xi }_{}}_{\min }=0, \\ 
 &  \nonumber\\ 
 & {{\xi }_{}}_{\max }=\frac{{{C}_{b }}\,V\cdot \left( {{\eta }_{h}}\,{P_{t,o}}\,{{C}_{0}}\,\beta  \right)}{\left( {{P}_{t,e}}\,{{T}_{b}} \right)} \nonumber 
\end{align}

The PDF of the end-to-end efficiency offers a mathematical model for predicting and analyzing its stochastic characteristics.

%% file: Sections/ResultsDiscussion.tex
\begin{table}[!t]
  \sisetup{group-minimum-digits = 4}
  \centering
  \caption{Cislunar Orbital Simulation Parameters \cite{donmez2026multi}}
  \label{tab:Kepler}
  \begin{tabular}{lllS[table-format=5]ll} 
    \toprule
    \toprule
     Altitude & 100 km  \\
     Eccentricity & $\approx0$   \\
     Inclination & $90^{\circ}$   \\
     True anomaly & $0^{\circ}$   \\
     Longitude of ascending node & $0^{\circ}$   \\
     Argument of perigee & $0^{\circ}$  \\
    \bottomrule
    \bottomrule
\end{tabular}
\vspace{-0.2 cm}
\end{table}

Table \ref{table:WPTSimParameters} shows the parameters used in power-beaming simulations. The optical transmitter power is almost halved during electro-optical conversion ($\eta_{eo}$) at a wavelength of 1064 nm. A small-sized solar panel area $A$ is selected since it is mounted on the transportable lunar vehicle and has a tracking feature. The horizon mask is a constraint that voids the link between the lunar receiver and the satellite laser when the angle between the lunar local horizon and the satellite transmitter is below 5$^\circ$. This threshold typically ranges between 5$^\circ$ and 15$^\circ$, and hence lunar obstacles can be avoided. 

 \begin{figure}[!t]
 \centering
\subfloat[]{
 	\label{subfig:7a}
 	\includegraphics[clip, width=0.95\columnwidth]{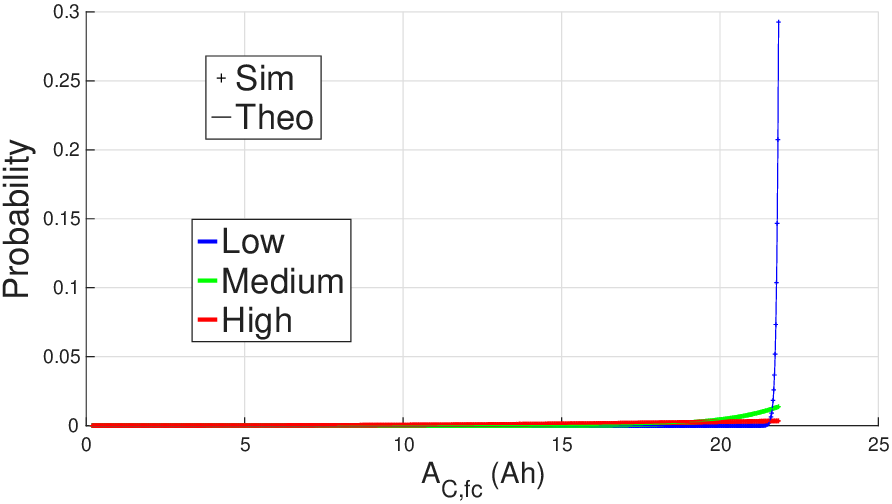}
 	 }
\\
\subfloat[]{
 	\label{subfig:7b}
 	\includegraphics[clip, width=0.95\columnwidth]{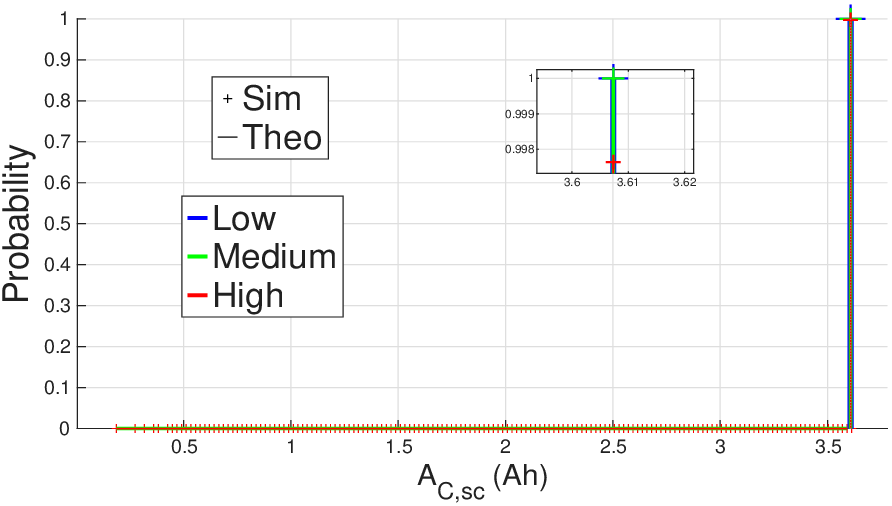}
 	 }
 \caption{Monte Carlo simulation versus theoretical PDF for low to high pointing errors: (a) fast charging $A_C$, (b) slow charging $A_C$.}
 \label{FIG:7}
 \end{figure}

Table \ref{tab:SpaceBatteries} shows the battery specifications used in the recharging-time calculations. The maximum $I_b$ is the most important factor because it determines how quickly a battery can charge fully. According to this table, the fast-charging and slow-charging modes offer different maximum $I_{b}$ values: 225 A and 30 A, respectively. The higher the $I_{b}$, the less time is required to fully charge the battery since the power caps are 832.5 W and 111 W for fast-charging and slow-charging modes, respectively.
 

The stochastic distribution of radial error angle, spatial displacement, and harvested power are established in the literature and in our work we incorporate realistic lunar scenarios for power beaming to a transportable-size solar panel mounted on a lunar vehicle. In addition to the realistic lunar power-beaming environment, we consider pointing errors with three different severities to understand the probabilistic distribution of $P_h$ for further analysis. Since a PDF is multiplied by a variable-specific bin width, our results present bin probabilities on the same scale for comparability. Unlike densities, bin probabilities do not exceed one. According to the results presented in Fig. \ref{FIG:5}, $v$ exceeds the solar panel radius only for the medium (moderate) and large (severe) pointing error levels, which can shift the beam footprint center outside the collector area.

In Fig. \ref{FIG:5} (a), the simulation matches the Rayleigh distribution for 0.5, 2.5, and 5 $\mu$rad upper error levels and hence the curves reach their maximum (i.e., $\sigma =\varphi_{c,max}/\sqrt3 $) at about 0.3, 1.3, and 2.9 $\mu$rad, respectively. According to the figure, the errors can exceed 16 $mu$rad which is more than three times the 5 $\mu$rad upper error level, since Gaussian tails are unbounded. In Fig. \ref{FIG:5} (b), interestingly, the peaks of the low and medium error curves fall inside 0.7 m of the solar panel radius, whereas the high one is significantly displaced and even reaches almost 7 meters. By leveraging the ratio $w(z)/b=6.32$, the solar cells can still collect optical power even when the displacement $v$ exceeds $b$. In Fig. \ref{FIG:5} (c), all curves reach around 631 W; however, the probability decreases with increasing errors. The high (5 $\mu$rad), medium (2.5 $\mu$rad), and low (0.5 $\mu$rad) error harvested power distributions have means of 629.46 W, 589.02 W, and 490.60 W, respectively. 

From Fig. \ref{FIG:6}--\ref{FIG:9}, our closed-form PDFs of random battery metrics that provide a rigorous analytical foundation are verified by extensive Monte Carlo simulations with $10\times10^6$ runs each. There are two charging modes, fast and slow charging and the mode transition parameter of $SoC_{FS}$ will be considered as a variable rather than a typical value of 80\% so we can compare pure results for the fast charging and slow charging to be able to make a fair comparison during the beaming time $T_b$.
 
In Fig. \ref{FIG:6} (a), $SoC_{fc}$ behaves similarly to $P_h$ as long as the 832.5 W power cap is not reached. The means of low, medium, and high $SoC_{fc}$ are 0.3453, 0.3360, and 0.3132, and hence the mean increases by 14.53, 13.60, and 11.32 percent over the initial 20\%, respectively. In Fig. \ref{FIG:6} (b), $SoC_{sc}$, a single spike at around 22.4\% can be seen for all error levels. The probability cap is around 0.99 for the high case, whereas this cap is 1.0 for the low and medium curves, and simulation and theory are consistent. This is because the power cap for slow charging is 111 W at 3.7 V, and $P_h\ge$111 W in 99.7\% of realizations for all error severities. It can be inferred that the mean $SoC_{fc}$ offers around 4.7 times the gain of the mean $SoC_{sc}$, even at high pointing-error levels.

 \begin{figure}[!t]
 \centering
\subfloat[]{
 	\label{subfig:8a}
 	\includegraphics[clip, width=0.95\columnwidth]{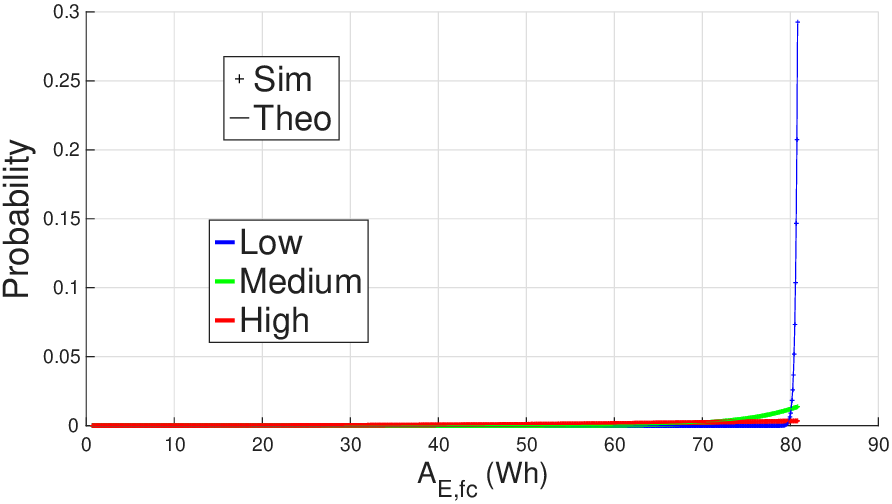}
 	 }
\\
\subfloat[]{
 	\label{subfig:8b}
 	\includegraphics[clip, width=0.95\columnwidth]{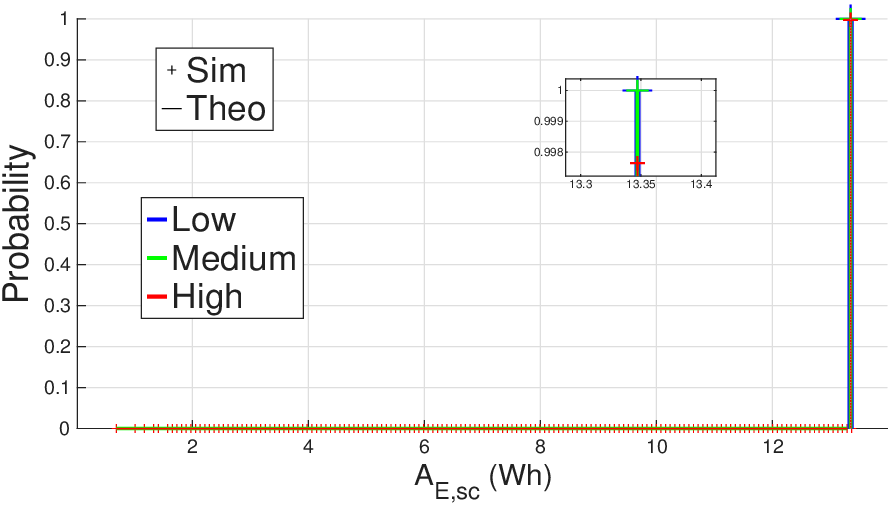}
 	 }
 \caption{Monte Carlo simulation versus theoretical PDF for low to high pointing errors: (a) fast charging $A_E$, (b) slow charging $A_E$.}
 \label{FIG:8}
 \end{figure}

In Fig. \ref{FIG:7} (a), accumulated capacity in both fast-charging and slow-charging modes is presented for various pointing error levels. The mean of $A_{C,fc}$ drops from 21.79 Ah to 16.99 Ah as the error increases from low to high. On the other hand, $A_{C,sc}$ collapses to a single point at 3.61 Ah for 0.99 and 1.0 for the high and other error levels, respectively, since the harvested current almost surely exceeds the 0.2 C current limit. Comparing the means, $A_{C,sc}$ can accumulate only one sixth of $A_{C,fc}$.

In Fig. \ref{FIG:8} (a), accumulated energy in both fast-charging and slow-charging modes is presented for various pointing error levels. The means of $A_{E,fc}$ are 62.85 Wh, 75.46 Wh, and 80.64h W for the high, medium, and low pointing error levels, respectively. On the other hand, $A_{E,sc}$ settles at approximately 13.34 Wh regardless of the error level. Therefore, the ratio of the means of $A_{C,fc}$ and $A_{C,sc}$ is about 5.66.

In Fig. \ref{FIG:9} (a), end-to-end efficiency in both fast-charging and slow-charging modes is presented for various pointing error levels. The mean values of $\xi_{fc}$ are 1.26, 1.18, and 0.98 percent for low, medium, and high pointing error severities, respectively. In contrast, $\xi_{sc}$ is pinned at approximately 0.21 percent with at least 0.99 probability for all pointing error levels, as the battery refuses power above 111 W. We can infer that slow charging mode is immune to the pointing errors, and this mode can utilize only around one-fifth of what the link can deliver even in severe pointing error.

%% file: Sections/Conclusion.tex
In our work, random pointing errors with three different severities are considered. The corresponding spatial error displacements can reduce the collected optical power when the displacement exceeds the radius of the circular solar panel mounted on the space vehicle. The distribution of random harvested power enables us to understand its stochastic behaviour; hence, system design parameters can be set accordingly in real-life applications, since STK astrodynamics simulations take important external forces into consideration.    

The closed-form PDFs of battery state-of-charge, accumulated capacity, accumulated energy, and end-to-end efficiency that we derive are validated by the extensive Monte Carlo simulations. These analyses are conducted for fast-only and slow-only battery charging modes for a fair comparison. We investigate whether the fast-charging mode outcomes can outperform slow-charging mode results significantly during the short beaming time, which is less than 10 minutes. The discrepancy between the two charging modes is quantified for various pointing error severity levels at the maximum target distance. We found that $\xi_{sc}$ is pinned at around 0.21 percent with at least 0.99 probability for all pointing error levels, as the battery refuses power above 111 W. On the other hand, the mean values of $\xi_{fc}$ are 1.26, 1.18, and 0.98 percent for low, medium, and high pointing error levels, respectively. Thus, $\xi_{sc}$ is immune to pointing errors, and this mode can utilize only around one-sixth of what the link can deliver in a mild pointing error.

 \begin{figure}[!t]
 \centering
\subfloat[]{
 	\label{subfig:9a}
 	\includegraphics[clip, width=0.95\columnwidth]{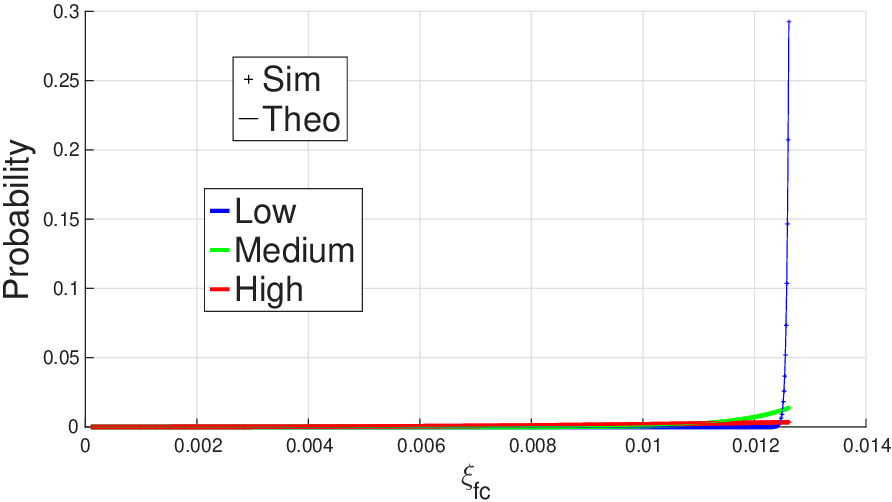}
 	 }
\\
\subfloat[]{
 	\label{subfig:9b}
 	\includegraphics[clip, width=0.95\columnwidth]{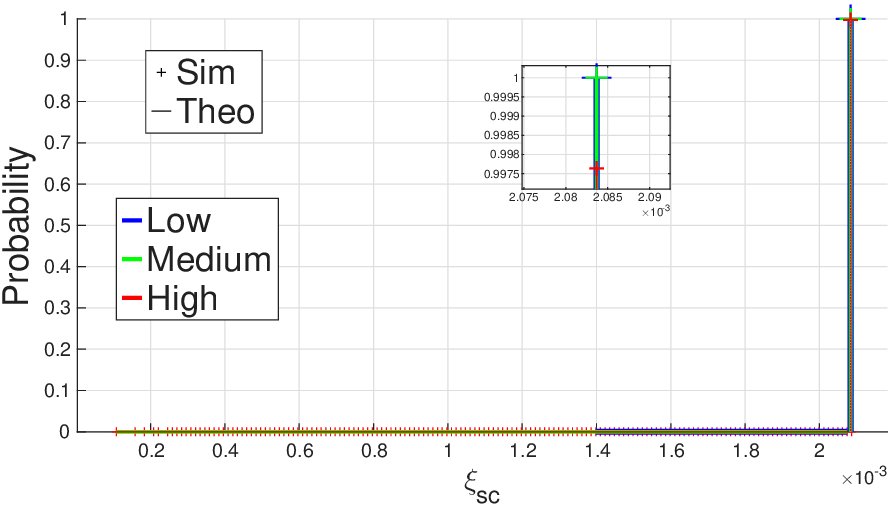}
 	 }
 \caption{Monte Carlo simulation versus theoretical PDF for low to high pointing errors: (a) fast charging $\xi$, (b) slow charging $\xi$.}
 \label{FIG:9}
 \end{figure}

%% file: Biography.tex
\begin{IEEEbiography}
[{\includegraphics[width=1in,height=1.25in,clip,keepaspectratio]{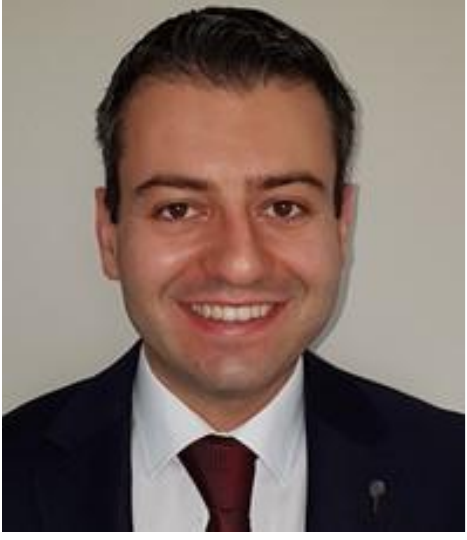}}] {Barış Dönmez}{\space}(Graduate Student Member, IEEE) received the  M.Sc. degree (with high honours) in electrical and electronics engineering in 2022 from FMV Işık University, Istanbul, Turkiye. He is currently pursuing his Ph.D. degree in electrical engineering at Polytechnique Montréal, Montréal, QC, Canada. His research interests include communication and energy harvesting systems in space networks.
\end{IEEEbiography}

\begin{IEEEbiography}
[{\includegraphics[width=1in,height=1.25in,clip,keepaspectratio]{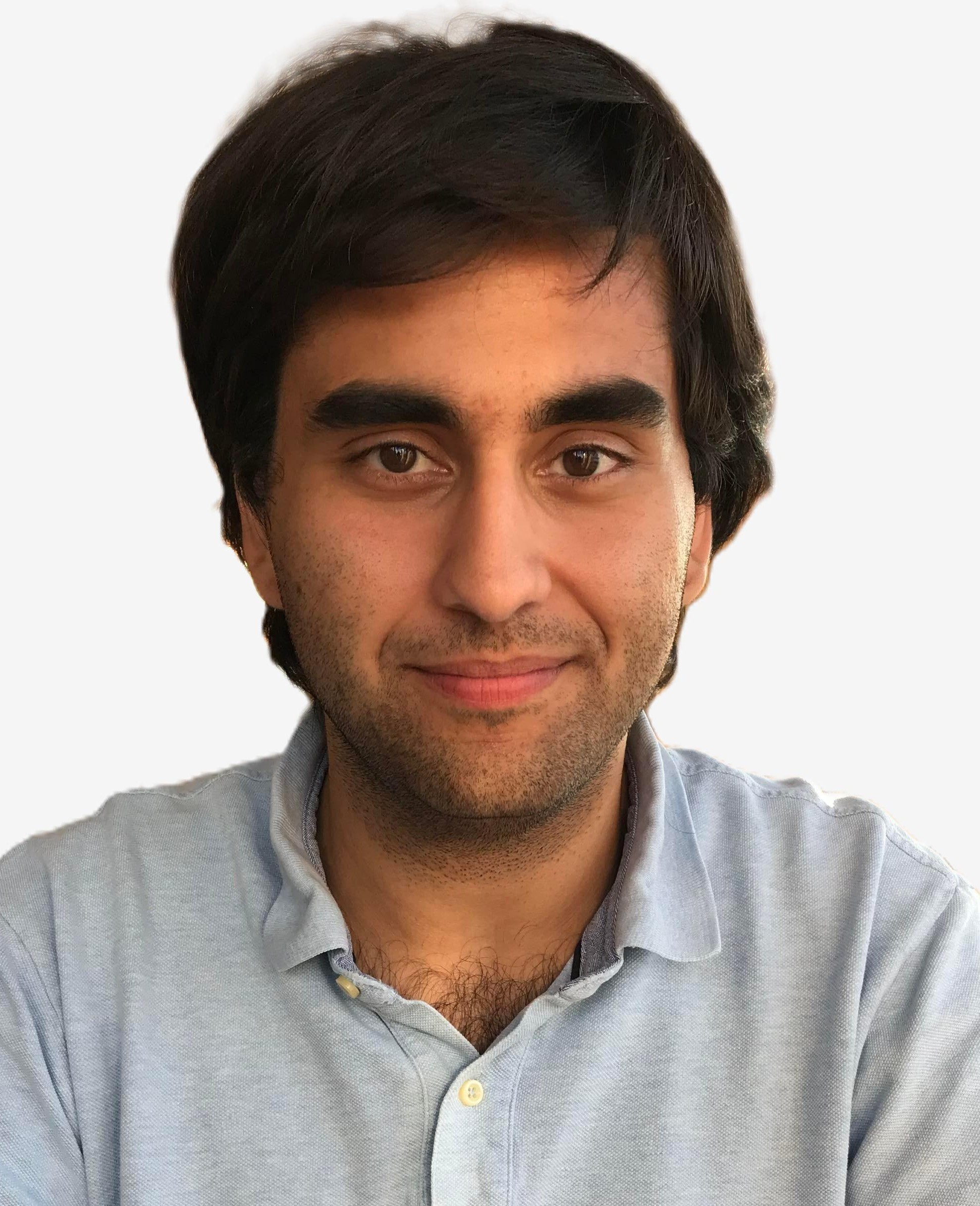}}] 
{Yanni Jiwan-Mercier}{\space} received his B. Eng. in physics engineering in 2020 and his M. Eng. in aerospace engineering in 2023, from École Polytechnique Montréal, Québec, Canada. He is currently doing his PH.D. degree in electrical engineering at Polytechnique Montréal on the subject of power transfer on the Moon. 
\end{IEEEbiography}

\begin{IEEEbiography}
[{\includegraphics[width=1in,height=1.25in,clip,keepaspectratio]{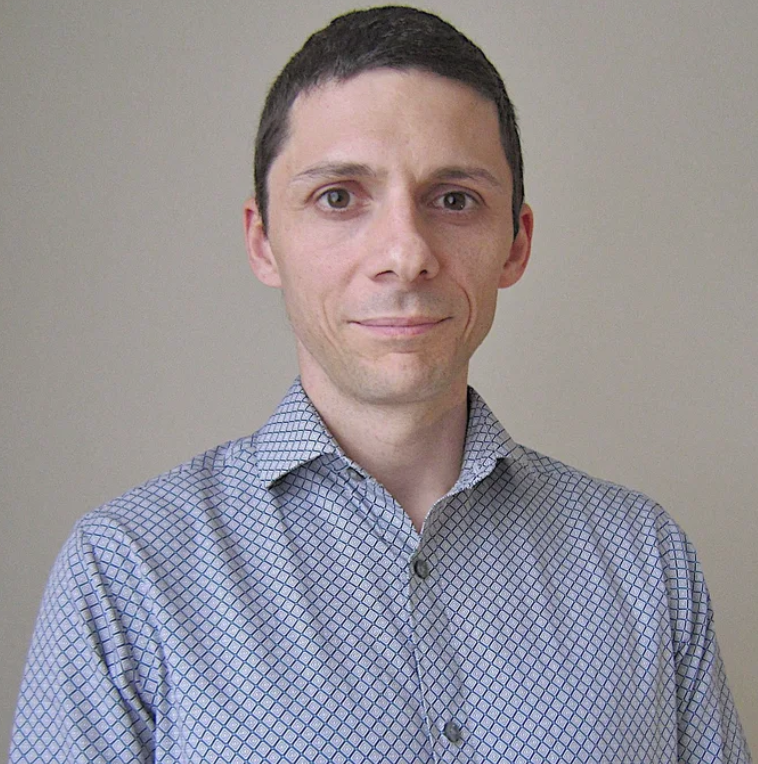}}] 
{Julien Magné}{\space} received the Ph.D. degree in electrical engineering from Université Laval, Québec, QC, Canada, in 2007, where he conducted his doctoral research at the Center for Optics, Photonics and Lasers (COPL) on pulsed fiber lasers and fiber Bragg gratings. He is currently Lead Laser Scientist at Volta Space Technologies in Québec, QC, Canada, where he works on laser power-beaming systems for lunar applications. His work spans both experimental development and modeling, including high-power fiber laser systems, beam pointing and tracking control, photovoltaic receivers for laser power conversion, and space hardware qualification. 
\end{IEEEbiography}

\begin{IEEEbiography}
[{\includegraphics[width=1in,height=1.25in,clip,keepaspectratio]{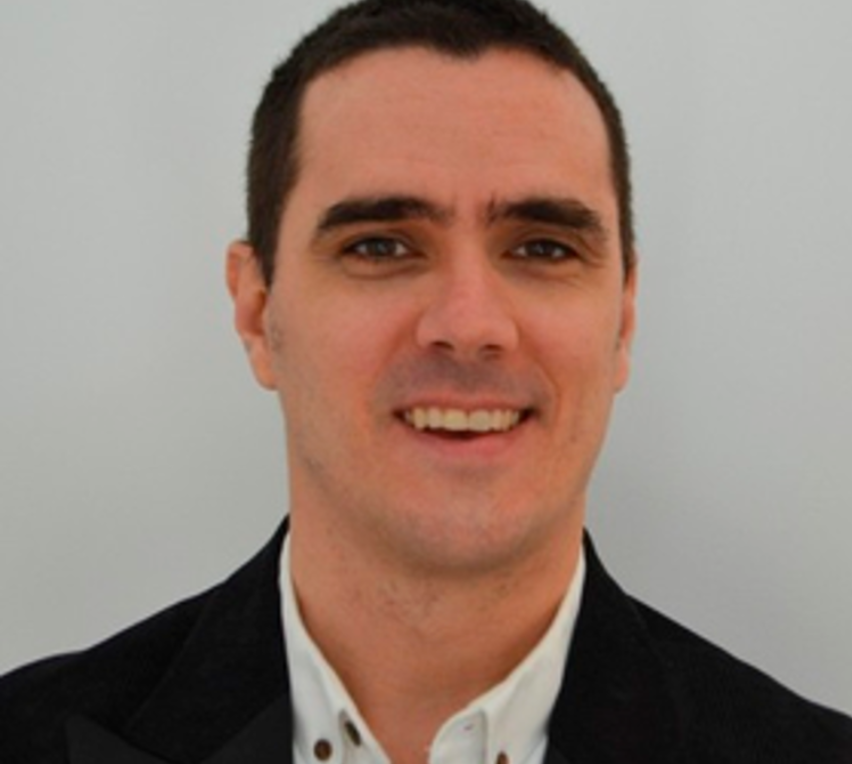}}] 
{Guillaume Blanchette}{\space} received his B. Eng. in physics engineering from École Polytechnique Montréal, Québec, Canada and holds multiple patents. He currently works as director of laser and technology development at Volta Space Technologies, where he leads the development of high-power spaceborne lasers and telescopes. He is an electro-optic expert with several years of hands-on experience developing high-power systems. He has several years of hands-on experience developing high-power ultra-fast lasers for multiple companies. He was also a product line manager at Electro Scientific Industries (Now MKS Instruments) and at Smiths Interconnect. At Smiths, he was instrumental in introducing and developing optical transceivers for sat-com and defence observatory satellite applications. 
\end{IEEEbiography}

\begin{IEEEbiography}
[{\includegraphics[width=1in,height=1.25in,clip,keepaspectratio]{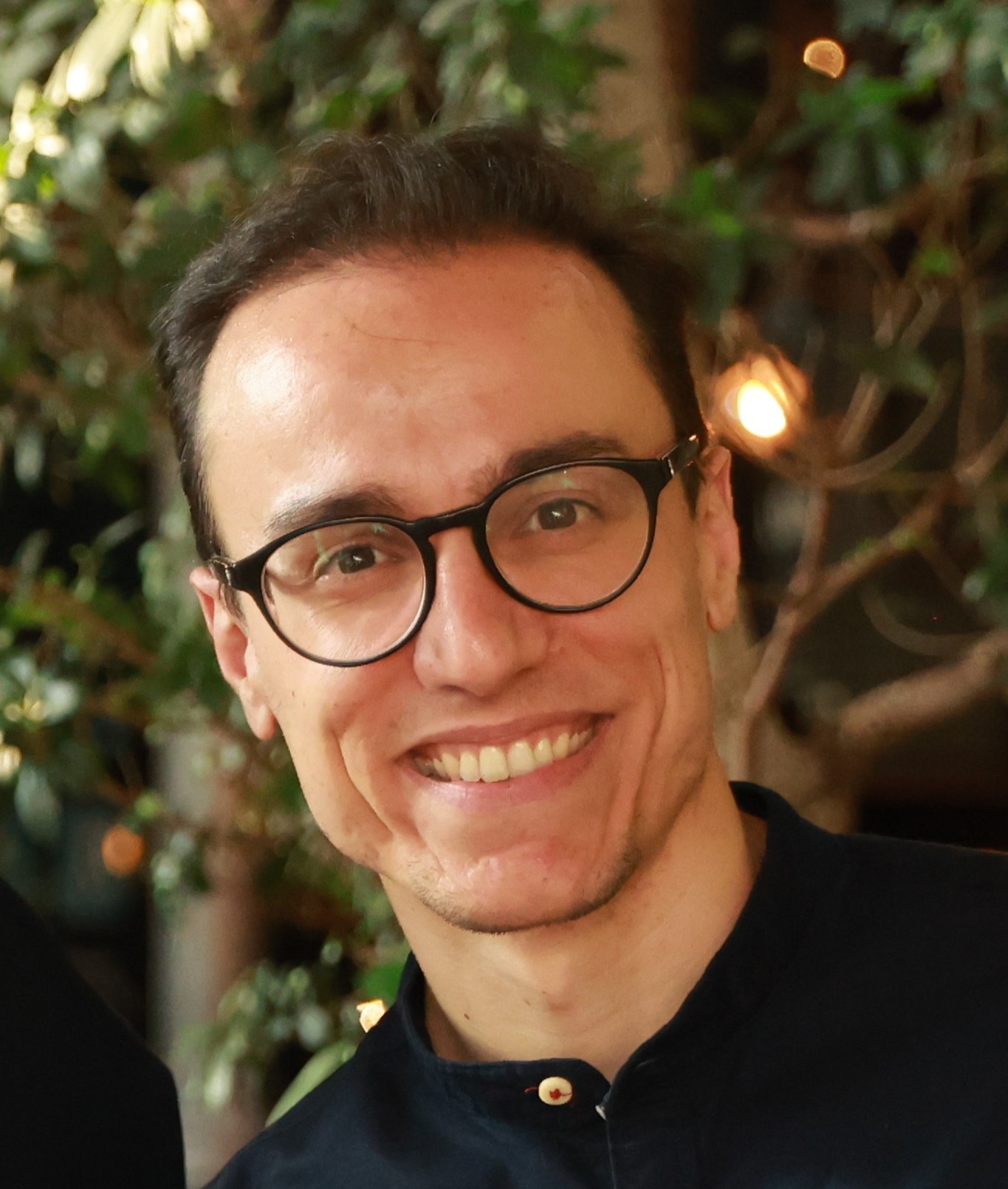}}] 
{Paolo Pino}{\space} received his Ph.D. in chemical engineering from Politecnico di Torino, Italy. He is the co-founder and Chief Technology Officer of Volta Space Technologies, where he leads the development of Volta's missions. He has a background in engineering and space exploration systems.
\end{IEEEbiography}

\begin{IEEEbiography}
[{\includegraphics[width=1in,height=1.25in,clip,keepaspectratio]{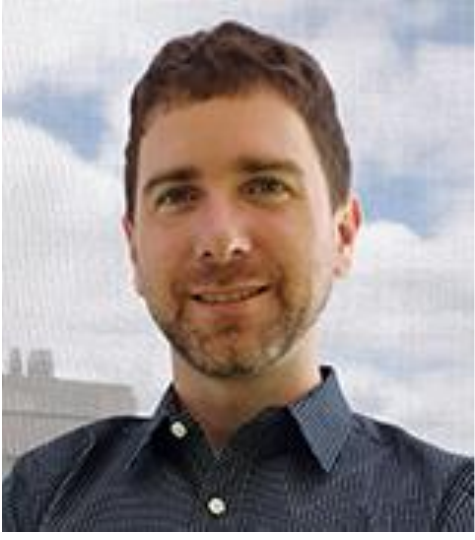}}] 
{Sébastien~Loranger}{\space}(Member, IEEE) is an Assistant Professor at Polytechnique Montréal, Montréal, QC, Canada. He received a Ph.D. in engineering physics from Polytechnique Montréal in 2018 and completed a postdoctoral fellowship at the Max Planck Institute for the Science of Light in Erlangen, Germany, in 2020. His research focuses on photonics and optoelectronic components for space systems.  
\end{IEEEbiography}

\begin{IEEEbiography}
[{\includegraphics[width=1in,height=1.25in,clip,keepaspectratio]{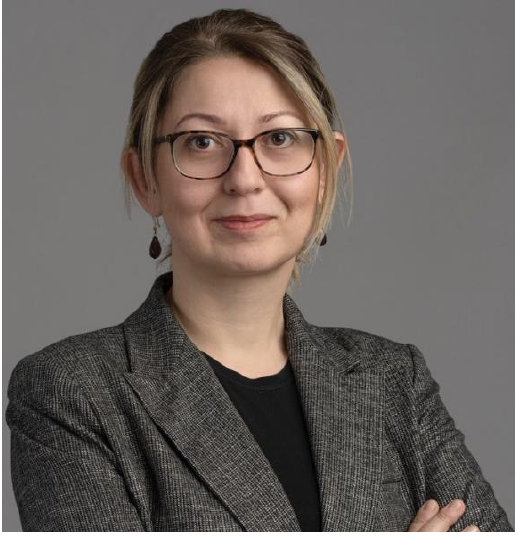}}] {Güneş Karabulut Kurt}{\space}(Senior Member, IEEE) received the B.S. degree from Bogazici University, Türkiye, in 2000, and the M.A.Sc. and Ph.D. degrees from the University of Ottawa, Canada, in 2002 and 2006. She is a Professor at Polytechnique Montréal, Canada Research Chair (Tier 1) in New Frontiers in Space Communications, Director of the Poly-Grames Research Center, and Co-Founder and Scientific Director of ASTROLITH. She is also an Adjunct Research Professor at Carleton University. From 2010 to 2021, she was a Professor at Istanbul Technical University. She is a Marie Curie Fellow and a Fellow of the Canadian Academy of Engineering.
\end{IEEEbiography}
